# An Intelligent Framework for Oversubscription Management in CPU-GPU Unified Memory

1st Xinjian Long, 2nd Xiangyang Gong
*State Key Lab. of Networking and Switching Tech.*
*Beijing University of Posts and Telecommunications*
Beijing, China
barbiel_origin@bupt.edu.cn, xygong@bupt.edu.cn

3rd Huiyang Zhou
*Department of Electrical and Computer Engineering*
*North Carolina State University*
Raleigh, USA
hzhou@ncsu.edu

*Abstract*—Unified Virtual Memory (UVM) improves GPU's programmability by enabling on-demand data movement between CPU memory and GPU memory. Thanks to this emerging feature, GPUs become more ubiquitous in systems ranging from servers to data centers, and there has been an increasing trend of adopting GPUs for large-scale and general-purpose applications. However, this trend soon creates a dilemma that the limited capacity of the GPU device memory is oversubscribed by the ever-growing application working set. Oversubscription overhead becomes a major performance bottleneck for data-intensive workloads running on GPU with UVM.

This paper proposes a novel intelligent framework for oversubscription management in CPU-GPU UVM. We analyze the current rule-based methods of GPU memory oversubscription with unified memory, and the current learning-based methods for other computer architectural components. We then identify the performance gap between the existing rule-based methods and the theoretical upper bound. We also identify the advantages of applying machine intelligence and the limitations of the existing learning-based methods. This paper proposes a novel intelligent framework for oversubscription management in CPU-GPU UVM. It consists of an access pattern classifier followed by a pattern-specific Transformer-based model using a novel loss function aiming for reducing page thrashing. A policy engine is designed to leverage the model's result to perform accurate page prefetching and pre-eviction. We evaluate our intelligent framework on a set of 11 memory-intensive benchmarks from popular benchmark suites. Our solution outperforms the state-of-the-art (SOTA) methods for oversubscription management, reducing the number of pages thrashed by 64.4% under 125% memory oversubscription compared to the baseline, while the SOTA method reduces the number of pages thrashed by 17.3%. Our solution achieves an average IPC improvement of 1.52X under 125% memory oversubscription, and our solution achieves an average IPC improvement of 3.66X under 150% memory oversubscription. Our solution outperforms the existing learning-based methods for page address prediction, improving top-1 accuracy by 6.45% (up to 41.2%) on average for a single GPGPU workload, improving top-1 accuracy by 10.2% (up to 30.2%) on average for multiple concurrent GPGPU workloads.

*Index Terms*—GPU, Unified Virtual Memory, Oversubscription, Deep learning

## I. INTRODUCTION

Modern GPUs support an advanced feature called Unified Virtual Memory (UVM) [1], which enables a unified virtual

Identify applicable funding agency here. If none, delete this.

memory space and programmer-agnostic demand-driven automatic data migration between the CPU memory and the GPU device memory. UVM significantly enhances GPU's programmability by relieving the developers from the burden of data management.

Despite the attractive benefits, UVM also raises concerns about the efficiency of the GPU runtime's data management strategy. Since there is only one physical copy of the data maintained in either the CPU memory or the GPU device memory, an inappropriate strategy may cause unnecessary memory transfers between the host and the GPU. As a result of the relatively slow bandwidth of the CPU-GPU interconnect, these transfers may lead to serious performance slowdown for the GPU workloads. Unfortunately, the limited capacity of the GPU device memory and the ever-growing size of the applications further increase the probability of the occurrence of this problem. Page thrashing unavoidably happens while the application's working set size exceeds the device memory size. Currently, page thrashing becomes a first-order performance bottleneck for data-intensive applications using GPU UVM.

A few methods have been proposed to handle page thrashing in GPU UVM. Ganguly et al. [2] studied the interplay between software prefetchers and page replacement algorithms under oversubscription. They proposed a new eviction policy inspired by the semantics of the tree-based prefetcher to reduce page thrashing. Furthermore, Ganguly et al. [3] leveraged the hardware-access counter and the zero-copy technique to propose an adaptive framework (**UVMSmart**) for page migration and pinning to address the performance overhead of oversubscribed GPGPU workloads. Yu et al. proposed a hierarchical page eviction policy [4] (**HPE**) for GPU UVM. These works utilize knowledge extracted from in-depth analysis of inherent application characteristics (memory access patterns, etc.) to design specific rules for each workload. Although these methods significantly mitigate the performance impact of memory oversubscription, there are still problems left to be solved. Firstly, a thorough understanding of workloads' memory access patterns in advance is not always possible in practice. A rule-based design for a particular subset of the GPGPU workloads may not be generic to the others. Secondly, memory access patterns may vary in different program phases. A simple combination of the existing data prefetchers and eviction policies can not handle this variance. Thirdly, existing mechanisms suffer from inefficiency when a data prefetcher and an eviction policy are combined.

To compensate for the weakness of the rule-based methods, machine intelligence is introduced to provide insights for improvement. Since there aren't any existing learning-based methods for oversubscription management in GPU UVM, we only discuss the learning-based works on other hardware prediction mechanisms in this paper. Hashemi et al. [5] apply the RNN model to the analysis of memory access patterns, which demonstrates higher precision and recall than table-based approaches. Shi et al. [6] applied deep learning to solve the cache replacement problem. Shi et al. [7] propose a hierarchical model of data prefetching that accommodates both delta patterns and addresses correlation. Existing learning-based works applied variant neural models and generated impressive performance improvement on different hardware issues by extracting knowledge from collected memory traces. However, unique challenges also emerge while these learning-based methods try to make themselves practical for use. Firstly, some of the basic operations of the learning-based methods (normalization, embedding, etc.) require profiling of future memory access information, which is not always available in practice. Secondly, as for the learning-based design which framed their issue as a classification problem, the number of classes may grow explosively in the life cycle of the running workloads. Such a growing number of classes may cause the neural models to suffer from serious forgetting problems when they are continually updated with new coming data. Thirdly, some of the existing learning-based methods use an identical neural model to handle all the data, and some of the works create neural models for every unique page. Empirically, it is difficult for a single model to learn the knowledge from all the memory access patterns or all the GPGPU workloads, whereas creating too many neural models may cost unacceptable computation and storage overhead.

To solve these challenges, we propose an intelligent framework for oversubscription management in CPU-GPU UVM in this paper. This framework takes as input a sequence of historical memory access information including page address, page delta, PC, and Thread Block ID. When the new data arrives, the input sequence will be first fed to a memory access pattern classifier to identify which pattern it belongs to. According to the pattern classification results, a specific set of neural model weights will be selected from a pattern-based model table and be fed to a novel page predictor. In other words, each pattern's input sequences are trained by a separate neural model. After being pre-processed, the input sequence will be fed to a novel page predictor to perform page delta prediction. To extract knowledge from both regular (stride, constant, etc.) and irregular (pointer chase, etc.) memory access patterns, the page predictor is composed of two Transformer-based [8] basic blocks to learn these patterns respectively. To solve the explosive growing number of classes problem, we introduce the incremental learning method into the page predictor to encourage the neural model to keep learning the new classes without forgetting the old ones. A novel loss function is used in the training of the page predictor, and this loss function helps the page predictor become thrashing-aware. All the prediction results with the same interval will be aggregated and be fed to a policy engine. By leveraging a prediction frequency table and a page set chain, the policy engine can learn the importance of different pages in the near future memory access of the workload. Then, the policy will determine the prefetching or the eviction candidates according to each page's importance when the corresponding prefetching or eviction request arrives. Finally, the decision will be sent to the GPU memory management unit (GMMU) and the corresponding memory operation will be performed.

This paper makes the following contributions:

- To our knowledge, this is the first paper to introduce a deep learning-based method into the oversubscription management in CPU-GPU UVM.
- We provide an in-depth analysis of the current rule-based methods for oversubscription management and the current learning-based methods on other computer architectural issues. We identify the necessity of applying machine intelligence for more accurate data prefetching and data eviction compared with the rule-based methods, and we identify sources of performance loss for current learning-based methods while running in an online manner.
- We propose an intelligent framework for memory oversubscription management in CPU-GPU UVM. Among 11 different GPGPU benchmark applications across different categories, our solution achieves a 64.4% reduction on average in page thrashing compared to the baseline under 125% UVM oversubscription, while the state-of-the-art (SOTA) work achieves a 17.3% reduction on average. Our solution achieves an average IPC improvement of 1.52X under 125% memory oversubscription, and our solution achieves an average IPC improvement of 3.66X under 150% memory oversubscription.
- Compared to the SOTA works which handle the input data online, our solution achieves a 6.45% top-1 accuracy improvement on average (41.2% at most) in page prediction of a single GPGPU workload, and our solution achieves a 10.2% top-1 accuracy improvement on average (30.2% at most) in page prediction of concurrent multiple GPGPU workloads.

The remainder of this paper is organized as follows. Section II presents the background of this work. Section III discusses the limitations of both the rule-based works on GPU memory oversubscription management and the learning-based works on other hardware prediction issues. Section IV describes the design of our intelligent framework. Section V compares the results of our intelligent framework with both the rule-based works and the learning-based works. Section VI discusses the related works of this paper, before providing concluding remarks in Section VII.

## II. BACKGROUND

In this section, we review the general mechanics of on-demand paging in CPU-GPU UVM, the soft and hard pining, the software and hardware prefetcher, and the page eviction mechanisms following the NVIDIA/CUDA terminology. It is worth noting that techniques mentioned in this section as well as our design described in the following sections are adaptable to the other GPU architectures besides NVIDIA.

### A. On-demand Page Migration and Soft/Hard Pinning

CPU-GPU UVM provides a single virtual address space accessible from both CPU and GPU. Using CUDA, developers can apply UVM by calling the **cudaMallocManaged** API to allocate data that can be accessed by both host code and GPU kernels with a single shared pointer. The functionality of Unified Memory is enabled by on-demand memory allocation and fault-driven page migration. In modern GPUs, load/store instructions use virtual addresses. When a scheduled thread/warp in an SM (Streaming Multiprocessor) generates a device memory access with virtual addresses, such virtual addresses are translated to physical ones before accessing data in the GPU L1 cache. The load/store unit (LDST) of that SM performs a translation lookaside buffer (TLB) lookup to find whether the translation for the issued memory access is cached in TLB or not. A miss in the last level TLB will be relayed to GPU memory management unit (GMMU), which performs a page table walk for the requested page. If there is a hit in either the TLB lookup or the page table walk, the translation will be returned and the requested data will be accessed within the GPU memory hierarchy. This is demonstrated as sequence (1) in Figure 1.

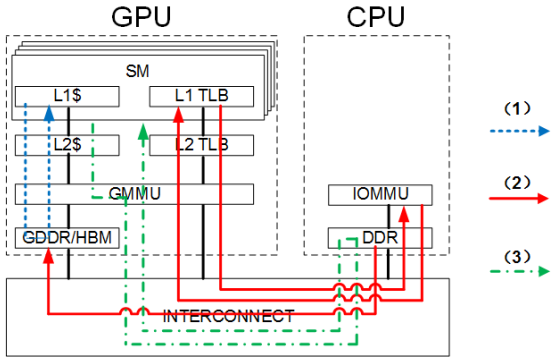

Fig. 1. Overview of the UVM page migration and zero-copy.

However, if there is no page table entry (PTE) for the requested page or the valid flag is not set, then a far-fault is registered in the GMMU's Far-fault Miss Status Handling Registers (MSHR) and the corresponding warp will be stalled. Then this request will be forwarded to the host and triggered a host-side page table walk. Once the page table walk is finished and the requested page is returned, MSHRs will be consulted to notify the corresponding LDST to replay the device memory access, and then the stalled warp will be marked executable.

This is demonstrated as sequence (2) in Figure 1, and this is the general process of GPU UVM on-demand page migration.

Handling far-faults with on-demand migration is costly because of the high latency of page table walk and data migration over PCI-e interconnect. The NVIDIA CUDA runtime introduces pinning memory to alleviate this problem. On one hand, developers can call the **cudaHostRegister** and the **cudaHostGetDevicePointer** APIs to force the memory allocation to be hard-pinned to the host memory. In this case, pages in such allocation of memory will never be transferred from host to device memory. GPU kernels can only request these pages using remote direct memory access (RDMA). This is demonstrated as sequence (3) in Figure 1, and this is the case of CUDA zero-copy. On the other hand, developers can call the **cudaMemAdviseSetAccessedBy** and the **cudaMemAdviseSetPreferredLocation** APIs to advise the allocation to be soft-pinned to the host memory. In this case, pages in such allocation will not be migrated to the device memory at the first touch. Rather, the migration will be delayed till the number of read-requests reaches a certain static threshold. This is demonstrated as the combination of (2) and (3) in Figure 1.

### B. Software and Hardware Prefetcher

CUDA 8.0 introduced the **cudaMemPrefetchAsync** API to handle the costly far-faults. This is a software prefetching scheme that allows the developers to manually overlap the kernel execution with the asynchronous data migration.

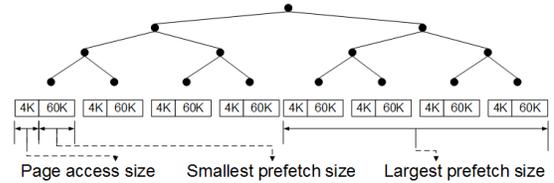

Fig. 2. A tree-based neighborhood hardware prefetcher implemented by NVIDIA since CUDA 8.0 on a 512KB memory chunk.

In the GPU Technology Conference 2018, a tree-based prefetcher was mentioned being implemented by NVIDIA CUDA 8.0 driver. Ganguly et al. [2] uncovered the semantic of this tree-based neighborhood prefetcher through micro-benchmarking and profiling. The user-requested size of a **cudaMallocManaged** allocation is logically divided into some $2MB$ memory chunks plus a remainder. Each of these chunk is further logically divided into $64KB$ basic blocks, which is the unit of prefetching. According to the far-faults received from the GPU, the runtime calculates the base addresses of the basic blocks corresponding to these faults. Then, these base addresses will be sent to the IOMMU and all the pages within the corresponding basic blocks will be migrated to the GPU. The runtime keeps track of the total size of valid memory resided in GPU for each non-leaf node among all the $2MB$ trees. If runtime detects that any non-leaf node's GPU valid memory is more than 50% of the total capacity of that node, the remaining non-valid pages of that node will be scheduled as further prefetching candidates. Figure 2 illustrates such a tree structure for a 512 $KB$ region.

*C. Page Eviction*

Plenty of research has targeted improving the performance of eviction policies for cache and memory on CPUs [9, 10, 11]. LFU is a representative frequency-based policy, but it is found not enough for selecting an appropriate eviction policy for unified memory [4]. The widely-used recency-based policy LRU performs well for a significant portion of applications. However, the ideal LRU is too expensive to implement, and it performs poorly for the thrashing access patterns. For unified memory, Zheng et al. [12] evaluated the performance of LRU and Random for some applications. LRU maintains an ordered list of pages based on their last access. Upon reaching GPU device memory capacity, LRU chooses the oldest accessed page. Unlike LRU, Random chooses a random page irrespective of when it is last accessed. Ganguly et al. [2] introduced a tree-based pre-eviction inspired by the tree-based prefetcher which leverages the full-binary tree structures created and maintained by the runtime for the managed allocations. It employs a thresholding-based heuristic inverse of the prefetcher. At any instance, if the current occupancy of a non-leaf node falls below 50%, the runtime pre-evicts other 64KB valid leaf nodes under it. Yu et al. [4] proposed a hierarchical eviction policy that manages a page set chain dynamically and uses statistics to classify applications. It selects an appropriate eviction strategy based on the classification result. GPU Technology Conference 2017 [13] specified that the CUDA drivers implement the LRU page replacement policy.

## III. CURRENT CHALLENGES

In this section, we describe the performance bottleneck associated with GPU UVM oversubscription. Then, we discuss the limitation of the existing rule-based methods for oversubscription management, and we discuss the limitation of the existing learning-based methods in computer architecture.

*A. Oversubscription Overhead*

Figure 3 shows that all GPGPU workloads suffer from performance loss due to memory oversubscription, and this loss exacerbates as the oversubscription level grows (the average performance slowdown under 125% oversubscription is 24.1%, and 3 of the applications (ATAX, NW, 2DCONV) are crashed under 150% oversubscription). More precisely, 125% oversubscription indicates that the device memory size equals 0.8 times the certain workload's working set size, and 150% oversubscription indicates that the device memory size equals 0.67 times the working set size. For instance, if the working set size of a certain application is 1 MB, then a device memory size of 0.8 MB will lead to 125% oversubscription. It is worth noting that the device memory size here only considers the managed memory allocated by calling **cudaMallocManaged**, and the **cudaMalloc** allocation are considered pinned and will not be evicted.

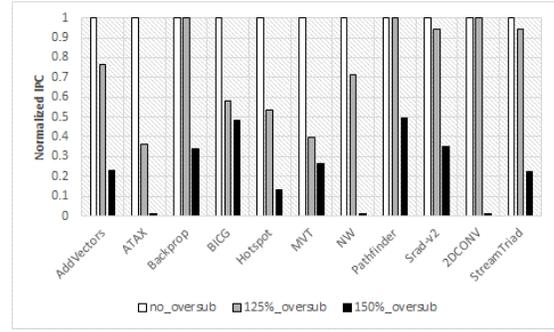

Fig. 3. Performance slowdown of GPU workloads under different percentages of memory oversbuscription.

The major cause of the performance loss of memory oversubscription (shown in Figure 3) is page thrashing, which means that pages are moved back and forth between CPU and GPU memory repeatedly upon reaching GPU device memory capacity. Since the data migration latency over PCI-e interconnect is very high due to the lower bandwidth compared to the local device memory (PCIe 3.0 interconnect's bandwidth is 16GB/s, AMD Instinct MI60 using HBM2's bandwidth is 1TB/s [14]), the higher the oversubscription level is, the larger the performance cost. In addition, the cooperation of inappropriate prefetching and pre-eviction mechanisms may further deteriorate this problem. While useless pages are fetched in by aggressive prefetching, useful pages may be evicted since the capacity of the device memory is reached, and these pages will have to be brought back in the near future. On the other hand, a recency-based eviction policy may mistakenly select the hot pages as eviction candidates because of the lack of information (frequency, pattern, etc.), which may cause instant thrashing and hurt the application performance. In the experiments are shown in Figure 3, the tree-based prefetching mechanism and the LRU pre-eviction policy are applied, which is the same as the CUDA runtime [1, 13].

*B. Limitation of existing rule-based methods*

Since page thrashing is the major performance bottleneck in UVM memory oversubscription, the number of pages thrashed becomes an important metric to evaluate the memory management strategy. Table I shows the number of pages thrashed using different strategies under 125% memory oversubscription. **Baseline** indicates the combination of tree-based prefetcher and the LRU eviction policy as described in Section III-A. Since Belady's MIN algorithm [15] (MIN) provably minimizes the number of cache misses, it is considered optimal guidance for a data replacement problem. We assume all the data fetched in are demand load (**D.** or **Demand.**) and MIN is applied for page eviction upon the capacity of the memory is reached. We take this combination (**D.+Belady.**) as the theoretical upper bound of this study. **HPE** [4] and **UVMSmart** [3] are the two SOTA works in handling the memory oversubscription in GPU unified memory.

TABLE I
TOTAL NUMBER OF PAGES THRASHED USING VARIANT MEMORY
MANAGEMENT STRATEGIES UNDER 125% MEMORY OVERSUBSCRIPTION.

| Benchmark | Baseline | D.+HPE | UVMSmart | D.+Belady. |
|---|---|---|---|---|
| AddVectors | 0 | 0 | 416 | 0 |
| ATAX | 4688 | 745 | 1728 | 0 |
| Backprop | 0 | 0 | 0 | 0 |
| BICG | 8704 | 8385 | 9952 | 2224 |
| Hotspot | 6144 | 0 | 4416 | 0 |
| MVT | 2912 | 0 | 2736 | 0 |
| NW | 29952 | 3230 | 23776 | 772 |
| Pathfinder | 0 | 0 | 160 | 0 |
| Srad-v2 | 5632 | 3942 | 5632 | 3667 |
| 2DCONV | 0 | 0 | 0 | 0 |
| StreamTriad | 0 | 0 | 0 | 0 |

We can see that **D.+Belady** achieves the minimum page thrashed for all the GPU workloads among all the strategies. However, since Belady's MIN algorithm is impractical, which requires future knowledge to decide which page to evict. This result can only be guidance instead of a solution to this problem. **Baseline** causes the largest number of pages thrashed for most of the workloads. We believe that the reason for this result is as described in Section III-A, which is because of the poor adaptability between the aggressive tree-based prefetching mechanism and the recency-based eviction policy. **UVMSmart** exploits both the page migration and the zero-copy to handle the memory oversubscription, which improves the performance of **baseline** to some extent. However, **UVMSmart** combines different prefetching and pre-eviction mechanisms according to the memory access pattern classification result in the profiling program phases. This combination is hard to keep effective when the memory access patterns in the following phases change significantly. Furthermore, the excessive use of pinned memory is risky and this may hurt the performance of the applications which use paged memory. **D.+HPE** achieves the second-best result besides with the upper bound. We believe that this is due to the demand load (no garbage prefetching) and the novel eviction policy (more information for eviction candidates' selection). After all, we can see that there is still space to improve between the SOTA works and the upper bound.

TABLE II
TOTAL NUMBER OF PAGES THRASHED USING DIFFERENT HPE-BASED
MEMORY STRATEGIES UNDER 125% MEMORY OVERSUBSCRIPTION.

| Benchmark | Demand.+HPE | Tree.+HPE |
|---|---|---|
| AddVectors | 0 | 377381 |
| ATAX | 745 | 498928 |
| Backprop | 0 | 14282372 |
| BICG | 8385 | 33797724 |
| Hotspot | 0 | 97340 |
| MVT | 0 | 0 |
| NW | 3230 | 8812785 |
| Pathfinder | 0 | 1878699 |
| Srad-v2 | 3942 | 44650411 |
| 2DCONV | 0 | 566653 |
| StreamTriad | 0 | 3690578 |

Table II shows that **HPE** experiences dramatic performance loss when it cooperates with data prefetcher to handle memory oversubscription (**Tree.** indicates the tree-based prefetcher). **HPE** relies on the per-page counter, which records the number of touched pages in each basic block to classify applications into regular and irregular access pattern types. This counter is substantially affected by data prefetching, and this makes it unable to deliver correct classification. It is worth noting that we only focus on the number of pages thrashed in this section, and demand load may provide more benefits compared to data prefetching on this metric. However, prefetching brings more performance improvement than demand load for the GPU workloads using UVM due to the slow CPU-GPU interconnect (described in Section II). This motivates the requirement of demand-load like prefetching, which means prefetching with high accuracy and coverage. And this raises the need for learning-based data prefetching mechanisms.

### C. Limitation of existing learning-based methods

Machine learning has provided insights into various computer architectural problems, including branch prediction [16, 17], cache replacement [6], and data prefetching [7, 18, 19, 20, 21]. It is natural to consider applying a learning-based method to memory oversubscription management. With the combination of an accurate prefetching mechanism and the corresponding eviction policy, we will be able to bring **D.+Belady.** (as described in Section III-B) into practicality. In other words, the more similar the prediction and the future knowledge of the access information are, the closer we are able to approach the upper bound in this problem.

Existing learning-based works [6, 7, 18, 19, 20, 21] can be formulated as follows:

$$P(Access_{t+1}|Access_1, Access_2, \cdots, Access_t) \quad (1)$$

The main idea of Equation 1 is to exploit correlations between consecutive historical memory access information to predict future memory access. In other words, a data prefetching problem can be viewed as a classification problem where each class is a potential indicator of a data location (address, address delta, offset, etc.) that will be accessed at time step $t + 1$ ($Access_{t+1}$), and the learning task is to learn the probability of these classes with the occurrence of all or a subset of historical memory accesses information ($Access_1, Access_2, \cdots, Access_t$). By leveraging machine learning, existing works show superior accuracy and coverage in data prefetching compared to the rule-based methods described in Section III-B. However, there are limitations to these designs. Previous works [7, 21] present online design by doing the training and prediction process in an alternate manner to accommodate the requirement of hardware prefetcher. Intuitively, this design makes sense since memory access patterns may vary in different program phases. The neural model needs to be retrained to capture the dynamic pattern, and then it can be used to deliver accurate predictions for data prefetching. However, we find that the reality may not the same as our intuition.

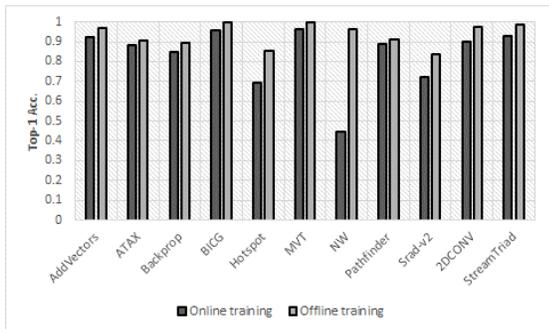

Fig. 4. Top-1 accuracy of page delta prediction on single workload using online training and offline training.

Figure 4 shows the results of page delta prediction on a single GPGPU workload using the online training method and offline training method. In the online training experiments, we use the starting 10% instructions (about 50 million instructions in most cases) of each GPGPU workload to train the model, and we use the trained model to make predictions for the following 10% instructions, and this train-predict loop continues until all the instructions are consumed. In the offline training experiments, we randomly select 50% of the total instructions as the training set to train the neural model, and then we use the trained model to make predictions for all the instructions in a temporal order. Both the online experiments and the offline experiments use the same input-output pair (input data is a sequence of 10 consecutive historical memory accesses, input features include address, address delta, and PC; output is the following memory access of the input sequence, output class is address delta), the same neural model (Transformer), and the same hyper-parameters. Figure 4 shows that there is an average 11.08% top-1 accuracy loss of page delta prediction using online training compared with using offline training. NW, Hotspot, and Srad-v2 are the three workloads that experience the largest accuracy loss (NW's accuracy loss is 51.4%, Hotspot's accuracy loss is 16.2%, Srad-v2's accuracy loss is 11.4%). We believe that the analysis of these workloads can reveal the reason for this performance gap between online training and offline training.

TABLE III
THE NUMBER OF UNIQUE PAGE DELTAS IN DIFFERENT PROGRAM PHASES.

| Bench mark | Program Phase | | |
|---|---|---|---|
| | 0 | 1 | 2 |
| AddVectors | 55 | 56 | 56 |
| ATAX | 112 | 113 | 114 |
| Backprop | 45 | 131 | 141 |
| BICG | 17 | 18 | 37 |
| Hotspot | 59 | 66 | 71 |
| MVT | 6 | 12 | 12 |
| NW | 479 | 830 | 1466 |
| Pathfinder | 98 | 102 | 103 |
| Srad-v2 | 49 | 145 | 170 |
| 2DCONV | 155 | 155 | 155 |
| StreamTriad | 38 | 38 | 38 |

Table III shows the number of unique page delta (the classification class used in the experiments described in Figure 4) in different program phases. We can see that NW's and Srad-v2's page delta numbers grow significantly from the beginning to the end. The distribution of NW's (Figure 5(a)) and Srad-v2's page delta (Figure 5(b)) also vary in different phrases. We believe that these are the major reasons for the accuracy loss of NW' and Srad-v2's page delta prediction using online methods: In each training batch, the neural model is updated with new coming training data whose number of classes is continually growing. Ideally, the neural model should be able to learn new patterns while maintaining the ability to recognize previous ones. However, machine learning or deep learning model always suffer from serious forgetting problems when they are continuously updated with new coming data. This is also termed 'catastrophic forgetting' problem [22].

According to Table III and Figure 5(c)-(d), Hotspot's and StreamTriad's page delta share a similar steady characteristic in both the quantity and the distribution. However, Stream-Triad's online prediction loss is smaller than Hotspot's. We reference the deterministic finite automaton (DFA) described in [3] to re-label all the input data with their corresponding access pattern. According to DFA's definition, there are 6 access patterns (described in Section IV-C) so we use the digits 0-5 to represent them. Figure 5(e)-(f) show the visualization of the access stream of StreamTriad and Hotspot workloads in different program phrases using these new labels. Compared to Hotspot, we can see that there is stronger temporal proximity of similar access patterns within StreamTriads' visualization. We believe that such pattern proximity helps the neural model perform better in a near local range within the application's global memory access stream. On the other hand, it is difficult for a single neural model to capture the knowledge in an online manner while the distribution of the memory access patterns is sparse. In this case, blindly increasing the training epoch of each batch may trigger the problem of over-training, which will cause even worse prediction performance in the subsequent batch. Figure 6 shows that it is better to use separate models to make predictions for different memory access patterns. As for Hotspot's page delta prediction, offline training delivers the highest top-1 accuracy (85.6%). Online training using multiple models delivers the second-best performance (80.5%) and online training using a single model experiences the largest performance loss (69.4%).

## IV. OUR SOLUTION

We describe our intelligent framework for oversubscription management in CPU-GPU UVM in this section. We start by presenting a high-level overview of the framework. We then describe the key innovations in our design. First, to handle the potential explosive growing number of classes in the new coming data, our framework uses incremental learning. Second, a novel neural model is used to perform data prefetching prediction with the awareness of thrashing information. Third, our framework adopts a pattern-aware scheme, so that instead of relying on a single model to handle all the data, our framework uses separate models to learn the knowledge from the data of different access patterns.

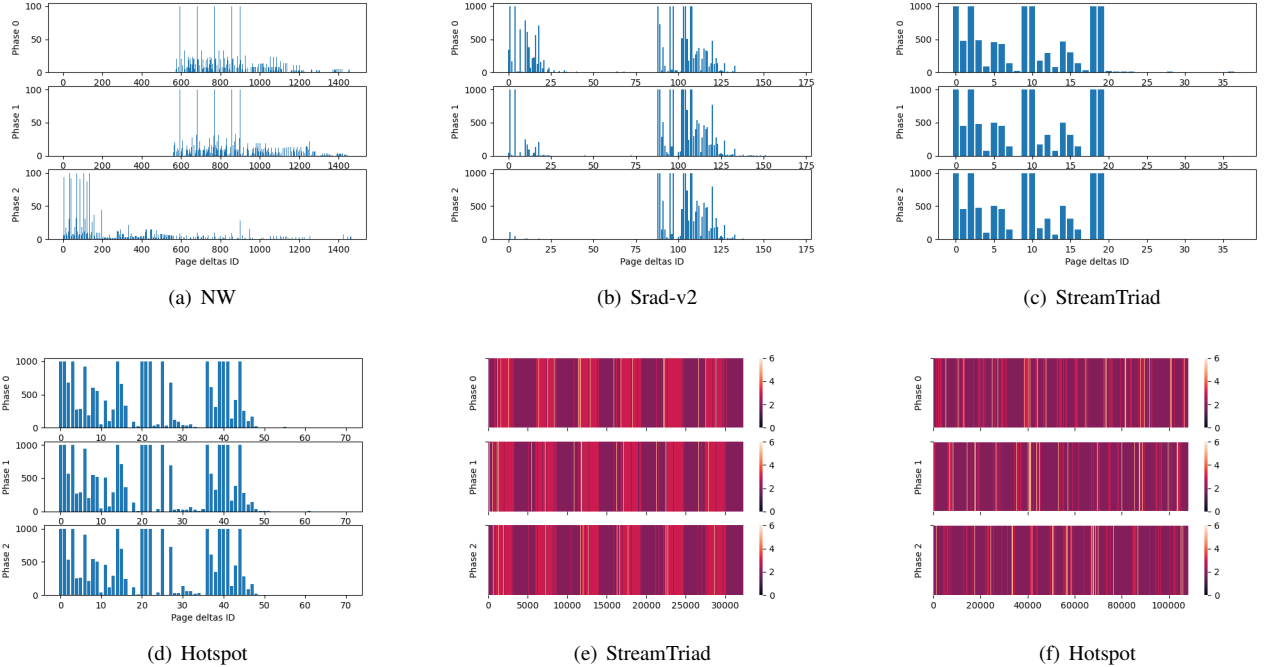

Fig. 5. Page delta distribution and memory access pattern visualization in different program phases.

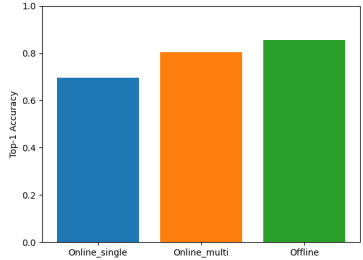

Fig. 6. Top-1 accuracy of page delta prediction of Hotspot using different training methods.

### A. Overall Design and Workflow

As shown in Figure 7, our intelligent framework takes as input a sequence of memory accesses and produces as output the next memory management policy to the GPU memory management unit (GMMU). The workflow of our framework is as follows: (1) The framework extracts features (page address, page address delta, PC, thread block ID) from the incoming memory access sequence; (2) According to the extracted features, there is a classification module to classify the sequence into different access-pattern categories; (3) Upon the classification result, a specific neural model corresponding to the certain pattern will be selected and be used to perform prediction; (4) To prepare the data for the prediction of the neural model, all the features will be transformed to numerical representation;

(5) Then, the selected neural model from (3) will apply the pre-processed input to perform prediction; (6) The policy engine decides the next memory strategies according to the prediction results and delivers it to GMMU; (7) GMMU receives the policy and performs the corresponding operations (including data prefetching, data pre-eviction, pinning, etc.).

### B. Thrashing-aware Incremental Learning based Page Predictor

Figure 8 shows our framework's neural model's architecture. In order to extract knowledge from both regular (data reuse, strides, etc.) and irregular (complex calculation with memory indirection, pointer chase, etc.) access patterns, the framework's neural model exploits four input features: page address, page address delta, PC, and Thread Block ID (TB ID). More precisely, address and delta are used to capture the regular access pattern, while PC and TB ID are used to capture the irregular access pattern. Thus, the two pairs of input features will be fed to two separate blocks (regular and irregular) for training after their individual embedding. Either block is an independent Transformer [8]. The reason for using Transformer is described in Section V-B. Finally, the outputs of these two blocks will be weighted by a learnable parameter respectively, and then they will be concatenated and be fed into a linear layer, producing a probability distribution over the classification classes. We use page delta as the output class in this study.

As described in Section III-C, we introduce incremental learning to handle the potential catastrophic forgetting problem caused by a continually growing number of classes over time. To avoid the knowledge acquired from the previous data being overridden by the new coming one, a regulation term is added to the model's loss function to consolidate previous knowledge when learning new data. Researchers have proposed different regulation-based methods to promote incremental learning performance. We use LUCIR [23] in this study. It is worth noting that our framework can also adapt to other more advanced regulation-based methods.

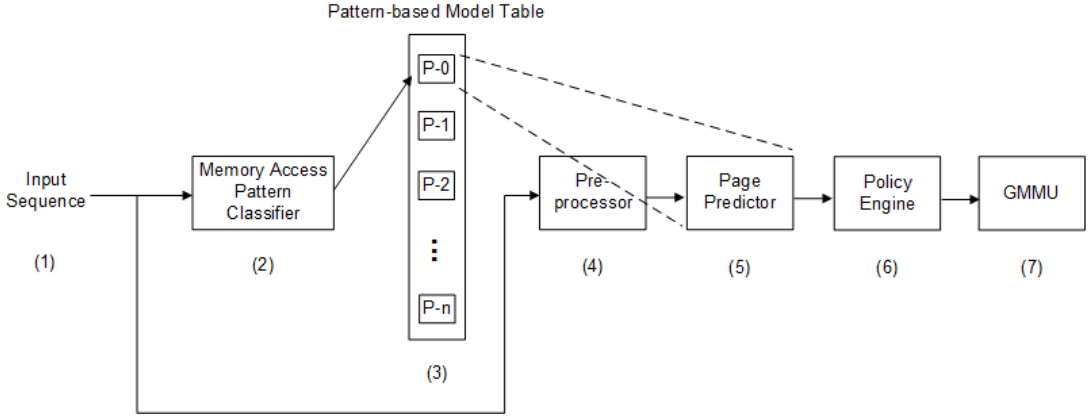

Fig. 7. Overview of intelligent framework for oversubscription management in CPU-GPU UVM.

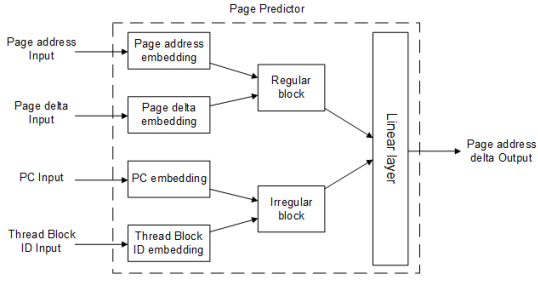

Fig. 8. Architecture of thrashing-aware incremental page predictor.

Since our final goal is to reduce oversubscription overhead, the page predictor should bring thrashing information into consideration, instead of only performing page prediction upon the correlations within the input sequence. We introduce a thrashing term into the page predictor's loss function, which is computed as follows:

$$L_{Thra}(x) = \sum_{i=1}^{|\mathcal{E}|\bigcup|\mathcal{T}|} y_i log(p_i) \quad (2)$$

$\mathcal{E}$ is the set of pages that have been evicted from the GPU device memory to the CPU memory, and $\mathcal{T}$ is the set of pages that have already been thrashed. $y$ is the one-hot ground-truth label and $p$ is the corresponding class probabilities obtained by softmax. Mathematically, $L_{Thra}(x)$ is the additive inverse of the standard cross-entropy (CE) loss. Semantically, $L_{Thra}(x)$ encourages the model to deliver page predictions beyond the set of evicted pages and thrashed pages, which aims to reduce the probability of pages being thrashed or being thrashed again.

Totally, the loss function $L$ of the page predictor is as follows:

$$L = \frac{1}{|\mathcal{N}|} \sum_{x\in\mathcal{N}} (L_{CE}(x) + \lambda L_{dis}^{G}(x)) + \frac{\mu}{|S|} \sum_{x\in|S|} L_{Thra}(x) \quad (3)$$

$\mathcal{N}$ is a training batch, which is composed of a set of pages corresponding to the newly arrived memory accesses. $\mathcal{S}$ is a subset of $\mathcal{N}$, which is defined as $|\mathcal{S}| = |\mathcal{N}|\bigcap(|\mathcal{E}|\bigcup|\mathcal{T}|)$. $L_{CE}(x)$ is the standard CE loss. $L_{dis}^{G}(x)$ is the regulation term introduced by LUCIR [23], which is used to encourage the orientation of features extracted by a current neural model to be similar to those by the previous model. $\lambda$ and $\mu$ are both loss weights. $\lambda$ is used to adjust the degree of need to preserve the previous knowledge according to the number of new classes introduced in each training batch. $\mu$ is used to adjust the degree of need to migrate the pages in the historical page thrashing and page eviction record according to the current memory access pattern. Intuitively, there are three goals that $L$ aims to achieve: Firstly, $L$ uses $L_{CE}(x)$ to perform multi-class classification learning upon the historical memory accesses. Secondly, $L$ uses $L_{dis}^{G}(x)$ to improve model training under the multi-class incremental setting. Thirdly, $L$ uses $L_{Thra}(x)$ to mitigate the probability of page prediction towards the ones which are already evicted or thrashed.

It is worth noting that replay-based methods [24, 25, 26] are also popular for alleviating the catastrophic forgetting problem. Replay-based methods store a set of reserved samples taken from all observed classes and replay the model on them to reduce forgetting. Commonly, there is a strict memory budget for the design of data prefetcher or cache replacement policy [27]. The implementation of the replayed-based methods may introduce dramatic storage overhead as the number of new classes grows. Thus, we do not consider the replay-based method in this study.

### C. Pattern-aware prediction scheme

As described in Section III-C, a memory access pattern classifier and a pattern-based model table are added to our intelligent framework (Figure 7) to help the neural model learn the knowledge of different access patterns. The pattern classifier takes the same input as the neural model (described in Section IV-B), and it determines which kind of pattern the input sequence of memory accesses belongs to. The pattern-based model table tracks the neural model of each access pattern. Since the architectures of all the neural models are identical, the pattern-based model table can be modeled as a direct-mapped cache that is indexed by a hash of the current access pattern and then returns the page predictor's weights for that pattern.

We reference DFA [3] as the pattern classifier in this study. DFA's mechanism is briefly described as follows. Based on the group of far-faults from GPU, unified memory runtime (UVM backend) determines the 64KB basic blocks as the migration candidates. These basic blocks are communicated to the I/O root complex to schedule DMA transfers. After segregating these transfers at kernel boundaries based on their scheduling timestamp, DFA scans the corresponding basic block addresses and determines whether they show linearity/randomness of migration. In addition, basic-block addresses are compared to determine any re-referencing across the kernel boundaries. Finally, the basic-block transfers are classified into 6 categories: Linear/Streaming, Random, Mixed/Irregular, Linear Reuse/Regular, Random Reuse, and Mixed Reuse. It is worth noting that our framework can also adapt to other more advanced pattern classification methods.

### D. Prediction-based memory strategy

Figure 9 shows the workflows of the policy engine (shown in Figure 7) handling page prefetching and page eviction.

When the GPU device memory fills to capacity, page eviction is required. In our intelligent framework, the eviction candidates are selected according to the page predictor's prediction results. To alleviate the probability of instant thrashing, we leverage the page set chain proposed by Yu et al. [4], which classifies the accessed pages into three partitions (new, middle, old) according to intervals (a specified number of page faults). We use this page set chain and update it with both the demand loads and the prefetches. The search for eviction candidates migrates from the old partition to the new partition depending on whether the certain partition is empty or not. In addition, we use a prediction frequency table to select the eviction candidate within the same partition. Typically, there is a large number of memory accesses within one interval, and there is equally a large amount of page prediction results for data prefetching in the same interval. The prediction frequency table keeps counters for pages occurring in several near intervals' prediction, which indicates the importance of these pages in the near future memory accesses of the workload. When a certain partition is chosen and the search begins, the eviction candidate will be selected from the pages with the lowest prediction frequency. As for pages that never show up in the prediction results, their frequencies are set to $-1$. This frequency table will be flushed periodically to keep an accurate representation of memory accesses in different program phases. We use the same interval length (64) as HPE, and we empirically flush the page frequency table every 3 intervals.

To deal with the long latency caused by on-demand paging, prefetching is required. As described in Section III-C and Section IV-B, the intelligent framework takes as input a sequence of historical demand loads' information (Page address, Deltas, PCs, and Thread Block IDs), and it generates as output a page delta between the current memory access and the next one. Data prefetching candidates are generated purely by prediction in each interval. An identical prediction frequency table as the eviction phase is used in data prefetching. This table can be exploited to control the amount of prefetching while the oversubscription level is too high. More precisely, the prefetching candidates will be selected from the pages with the highest prediction frequency. In this study, we use all the predicted pages as prefetches. The length of the historical sequence is empirically set to 10.

### E. Hardware Complexity

TABLE IV
MEMORY FOOTPRINT USING PATTERN-AWARE PREDICTION SCHEME (WITH QUANTIZATION).

| Benchmark | Params.(MB) | Acti.(MB) | Patterns | Total(MB) |
|---|---|---|---|---|
| AddVectors | 0.41 | 1.46 | 3 | 6.84 |
| ATAX | 0.27 | 1.46 | 3 | 6.00 |
| Backprop | 0.73 | 1.46 | 3 | 8.76 |
| BICG | 0.71 | 1.46 | 3 | 8.64 |
| Hotspot | 0.50 | 1.46 | 3 | 7.38 |
| MVT | 0.50 | 1.45 | 3 | 7.35 |
| NW | 0.48 | 1.47 | 4 | 9.72 |
| Pathfinder | 0.57 | 1.46 | 3 | 7.80 |
| Srad-v2 | 0.50 | 1.46 | 3 | 7.38 |
| 2DCONV | 0.46 | 1.46 | 3 | 7.14 |
| StreamTriad | 0.42 | 1.46 | 3 | 6.90 |

Table IV shows the memory consumption of the pattern-aware prediction scheme. These memory consumption statistics are collected using a MIT-licensed library [28]. Equation 4 shows the calculation of the total memory footprint.

$$Total = (Params. \times 2 + Acti.) \times Patterns \quad (4)$$

$(Params. \times 2 + Acti.)$ indicates that both the weights of the current and the previous models need to be stored for the calculation of the LUCIR term, but only the current model's activation is needed to update the model's weights. This sum is multiplied by $Patterns$ because each access pattern needs an individual model to make predictions. We exploit quantization to compress the memory consumption of our solution. Explorations show that clamping the weights and the forward/backward pass activation to [-16,+16] will not harm the performance of our predictor. We believe that the memory footprint of our solution could be mitigated compared to the one using the float32 value (5 bit is enough to represent all the values in our predictor). Overall, the largest storage cost of the pattern-aware prediction scheme is 9.72MB, which is significantly smaller than a state-of-the-art deep-learning-empowered data prefetcher [7, 29]. In addition, a new Transformer engine using FP16 precision has been built upon the latest NVIDIA Hopper Tensor Core [30]. We believe that the hardware complexity of the pattern-aware prediction scheme can be further improved in the latest GPU architecture.

The prediction frequency table is designed as a 16-way set associate cache (which is similar to the shared GPU L2 cache). Each entry is corresponding to a basic block (described in Section II-B), which means the tag is 48 bits (system width 64 bits, page address width 12 bits, basic block address width 16 bits). The data field of each entry is used to store the counter of the predicted pages within each basic block. Since we flush this table every 3 intervals, a 6-bit counter for each page is adequate in our experiments. In addition, the length of this frequency table is set to 1024 because the largest working set in our experiments is 64MB (a larger working set is impractical for simulation due to the long-running time). Thus, the total storage overhead of the prediction frequency table is 18KB ((6*16+48)/8*1024=18KB), which is relatively small compared to the pattern-aware prediction scheme.

TABLE V
CONFIGURATION PARAMETERS OF GPGPU-SIM.

| Simulator | GPGPU-Sim UVM Smart |
|---|---|
| GPU Architecture | NVIDIA GeForceGTX 1080Ti Pascal-like |
| GPU Cores | 28 SMs, 128 cores each @ 1481 MHz |
| Shader Core Config | Max 32 CTAs and 64 warps per SM, 32 threads per warp GTO scheduler |
| Page Size | 4KB |
| Page Table Walk Latency | 100 core cycles |
| CPU-GPU Interconnect | PCI-e 3.0 16x, 8 GTPS per channel per direction, 100 GPU core cycles latency |
| DRAM Latency | 100 GPU core cycles |
| Zero-copy Latency | 200 GPU core cycles |
| Far-fault Latency | $45\mu s$ |

similar to [7]). For each GPGPU workload using the offline methods, we randomly select 50% of their total instructions as the training set to train a single neural model, and then we use the trained model to make predictions for all the instructions in temporal order (this method is similar to [6]).

To compare with different memory strategies under oversubscription, we use a GPGPU-Sim extension implemented by Ganguly et al. [3] in our experiments. This extension provides functional and timing simulation support for UVM. Furthermore, this extension supports a smart runtime, which is composed of (1) a detection engine to identify the pattern in CPU-GPU interconnect traffic, (2) a dynamic policy engine that chooses from a wide array of existing memory management policies, and (3) an augmented memory management module that adaptively switches between delayed page migration and pinning. This extension also provides a set of regular and irregular GPU applications from Rodinia, Lonestar, and Polybench benchmark suites. These benchmarks are modified to use CUDA UVM APIs (**cudaMallocManaged**, **cudaMemPrefetchAsync**, and **cudaDeviceSynchronize**). Table V shows the primary configuration of the simulator, and the configuration associated with the UVM Smart runtime is the same as in [3].

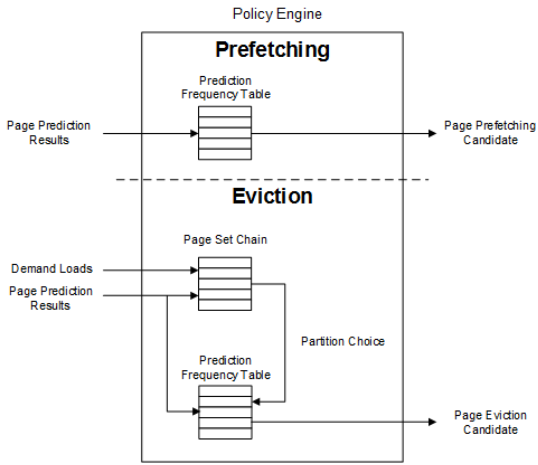

Fig. 9. Overview of intelligent framework's policy engine.

## V. EVALUATION

We evaluate our framework from two aspects. Firstly, we compare prediction performance running with the previous training methods and our thrashing-aware incremental-learning-based page predictor. Secondly, we compare prediction overhead sensitivity, IPC, reduction in thrashing, and scalability running with our framework and a state-of-the-art framework (UVMSmart) [3] which supports delayed page migration, zero-copy, and tree-based page prefetching.

### A. Evaluation Methodology

To compare with different learning-based methods, we implement different model-based predictors in Pytorch. For each GPGPU workload using the online methods, a single neural model is trained with a group of 50 million instructions, and we use this trained model to make predictions for the subsequent group of 50 million instructions. And we keep this routine until all the instructions are consumed (this method is

In order to hide the long latency of model training, we randomly select 5 benchmark applications (ATAX, Backprop, BICG, Hotspot, NW) and run them using different input data sets compared to the simulations described in Section V. We use 50% of each of these benchmarks' simulation results to build a corpus, and we train our predictor described in Section IV-B on this corpus until its accuracy reaches a reasonable range ($\geq 0.85$ in our experiments). We use this pre-trained model to make predictions for each benchmark, and we fine-tuned this model in each simulation every 50 million instructions to make it become adaptive in different program phases. According to our statistics among 11 benchmarks, this training method introduces a microsecond-level inference overhead for each prediction. According to NVIDIA's announcement [31], the inference latency of BERT-large (with 345 million parameters) could be slashed to 1.2 $ms$ by leveraging the TensorRT 8.0 SDK. In addition, a new Transformer engine has been built upon the latest NVIDIA Hopper tensor core [30]. This engine uses FP8 and FP16 precisions to reduce memory storage and increase performance. Since our solution is also built upon the Transformer model (Section IV-B) using a smaller precision than FP8 (Section IV-C), both the hardware overhead and the prediction overhead of adopting our solution are ideally negligible for the latest GPU architecture. Shi et al. [7] also claim that their model (which is composed of LSTMs and Transformers with a much larger model size than our predictor) can make predictions every 18000 nanoseconds. We believe that the prediction overhead can be improved by more advanced hardware/software technologies, more fancy equipment, and more sophisticated programming skills. However, these are out of the scope of this study. We conduct a prediction overhead sensitivity test of our predictor (described in SectionV-C), and we select one of the candidate overheads to perform the simulation in the subsequent evaluation.

### B. Prediction Performance

As described in Section VI-B, researchers have adopted different AI-based approaches to divergent computer architectural problems. More precisely, some of these works [7, 18, 19, 20, 21, 32, 33, 34, 35] are targeting data prefetching problems (beyond GPU UVM). Figure 10 shows the comparison results of using different predictors to deliver page delta prediction using online training. We can see that the Transformer-based method delivers the best prediction performance compared to the other methods (Convolution Neural Network, LSTM, Multi-Layer Perceptron). Thus, we select Transformer as the regular and irregular blocks' neural model in our framework.

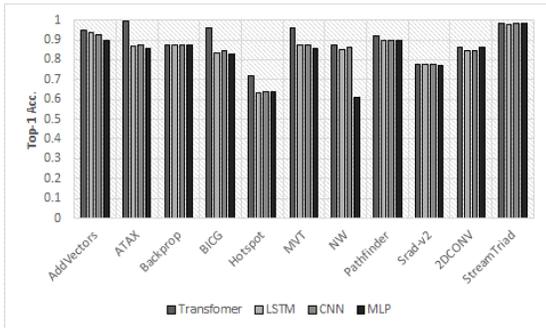

Fig. 10. Prediction performance using different page predictors.

Since offline training (profiling-based training) exploits future memory access information for training. We take offline training performance as the upper bound of the page prediction, and Figure 11 shows the normalized top-1 accuracy of both online training and our solution according to offline training (described in Section V-A). Different workloads experience different levels of performance loss when the input data arrives in temporal order (described in Section III-C). Our solution alleviates this performance loss by introducing incremental learning (Section IV-B) and pattern-aware prediction scheme (Section IV-C) into the UVM page prediction. Compared to the online training method, our solution achieves a 6.45% top-1 accuracy improvement on average (41.2% at most) in page delta prediction on a single GPGPU workload.

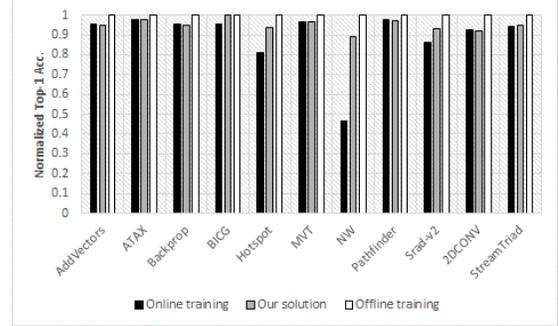

Fig. 11. Normalized top-1 accuracy of page delta prediction on a single workload using online training, offline training, and our solution.

Figure 12 shows the prediction performance of the loss function with or without the thrashing term (described in Section IV-B). We select 4 applications (ATAX, BICG, NW, Srad-v2) that experience the most serious page thrashing. By leveraging the thrashing term (w. term), our solution achieves an average 7.4% page thrashing reduction while causing a minimal top-1 accuracy loss (1.2%) compared to the one using only LUCIR loss (w/o. term). In our experiments, the loss weight $\mu$ is adaptively adjusted ranging from (0, 1]. This variance is caused by the applications' different memory access patterns. For instance, a larger thrashing term value benefits for the streaming access pattern where pages are hardly re-referenced once they are accessed. On the other hand, a smaller thrashing term value benefits the most repetitive access pattern where pages are referenced multiple times with different frequencies.

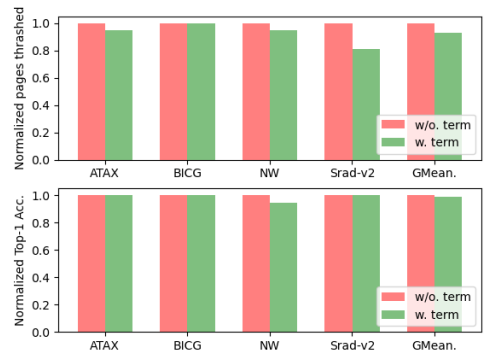

Fig. 12. Performance of the loss function with or without the thrashing term.

### C. Prediction Overhead Sensitivity Tests

Figure 13 shows the normalized IPC results of the prediction overhead sensitivity tests using 11 GPU benchmarks under 125% memory oversubscription. We vary the inference latency from 1, 10, 20, 50, and 100 microseconds per prediction.

Since the GPU core frequency is configured as 1481 MHz in the simulator, these latency candidates roughly correspond to 1500, 15000, 30000, 75000, 150000 cycle-per-prediction in each simulation. We consider UVMSmart as a state-of-the-art (SOTA) design. The average normalized IPC results under different levels of prediction overheads are 1.52X (1 microsecond), 1.32X (10 microseconds), 1.17X (20 microseconds), 0.91X (50 microseconds), and 0.71x (100 microseconds). Compared to the SOTA design, our predictor can achieve a 50% average IPC improvement when the prediction overhead is 1 microsecond, but this improvement vanishes and becomes a 10% performance slowdown when the overhead grows to 50 microseconds. Such deterioration continues and turns into a 30% performance slowdown when the overhead grows to 100 microseconds. These results show that our predictor, as well as other learning-based methods, are sensitive to the prediction overhead. In our subsequent experiments, we assume that our framework is situated at the UVM backend to make predictions. We use 1 microsecond (1500 cycle) as the prediction overhead, which is sharply distinct from the previous works [6, 7] that consider zero prediction overhead while exploiting deep learning models to boost the application's IPC performance. The training overhead of the pre-trained model is not considered in the simulation, we assume that it can be achieved offline in practice.

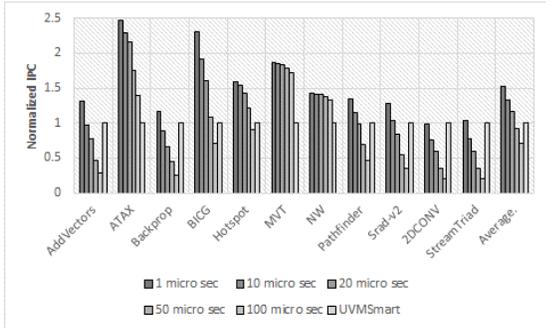

Fig. 13. Normalized IPC of 11 GPU benchmark applications using the revised predictor with different levels of prediction overhead under 125% memory oversubscription.

### D. IPC

Figure 14 shows the normalized IPC of UVMSmart and our solution across regular and irregular applications under different memory oversubscription levels. When the memory oversubscription level is 125%, our solution achieves performance improvement for all the benchmarks ranging from 3% to 140%. When the memory oversubscription level is 150%, some of the benchmarks (ATAX, NW, 2DCONV) using UVMSmart runtime crashed due to serious page thrashing. Thanks to the power of deep learning, our solution prevents these benchmarks from crashing through accurate prefetching and pre-eviction, and our solution achieves performance improvement for most of the benchmarks ranging from 131% to 328%.

Overall, our solution improves IPC by an average of 52.53% (geometric mean) compared to UVMSmart under 125% memory oversubscription, and our solution improves IPC by an average of 266.65% (geometric mean) compared to UVMSmart under 150% memory oversubscription.

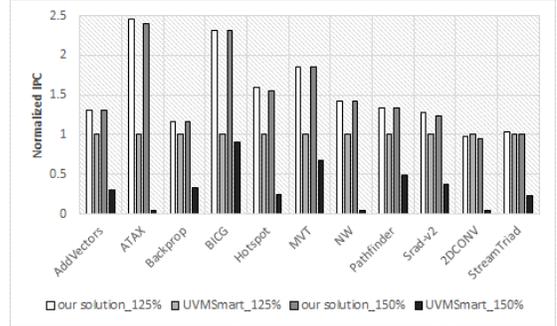

Fig. 14. Normalized IPC results using variant memory management strategies under 125% and 150% memory oversubscription.

### E. Reduction in thrashing

Table VI shows that our solution achieves the least number of pages thrashed compared to other methods (**Baseline**, **Tree.+HPE**, **UVMSmart**) involving both data prefetching and eviction. Thanks to the more accurate page prefetching using the learning-based method, the aggressiveness of the rule-based prefetcher (e.g., tree-based prefetcher) are moderated. The cooperation between the prefetching and the eviction policy is also improved by applying a thrashing-aware neural model and a shared data structure (Section IV-D). There is still space to improve between our solution and the methods (**Demand.+HPE**, **Demand.+Belady**) involving demand loads and eviction policy. This result indicates that our solution can not deliver perfect prediction, and there are still useless prefetches within our page prediction that cause performance loss compared to demand loads. It is worth noting that demand load is not practical for workloads using GPUs with unified memory (as discussed in Section II and Section III-B). Without considering **Tree.+HPE** (which is malfunctioning due to the poor cooperation between the prefetching and the eviction policy) and the demand loads-based methods (**Demand.+HPE**, **Demand.+Belady**), our solution achieves a 64.4% reduction on average in pages thrashed compared to the baseline under 125% memory oversubscription, while the SOTA works (**UVMSmart**) achieves a 17.3% reduction on average compared to the baseline. Page reduction under 150% memory oversubscription is not discussed in this section because some of the benchmarks using UVMSmart are crashed (Section V-D).

### F. Scalability

Modern GPUs allow multiple kernels or applications to share a single device concurrently [36]. This feature raises the challenge for the intelligent framework, as well as the page predictor, in handling memory oversubscription with an even larger amount of newly arriving classes and more complicated memory access patterns (described in Section III-C) compared with running with a single workload's input. We test the scalability of our solution by running multiple workloads belonging to different categories (streaming, regular, mixed, random) concurrently. Table VII (S.T. indicates StreamTriad, Hot. indicates Hotspot) shows the results of these experiments. Thanks to the incremental-learning-based predictor and the framework's pattern-awareness, our solution achieves a 10.2% top-1 accuracy improvement on average (30.2% at most) in page delta prediction on multiple GPGPU workloads.

TABLE VI
TOTAL NUMBER OF PAGES THRASHED USING VARIANT MEMORY MANAGEMENT STRATEGIES UNDER 125% MEMORY OVERSUBSCRIPTION.

| Benchmark | with Prefetching | | | | without Prefetching | |
|---|---|---|---|---|---|---|
| | Baseline | Tree.+HPE | UVMSmart | Our solution | Demand.+HPE | Demand.+Belady. |
| AddVectors | 0 | 377381 | 416 | 60 | 0 | 0 |
| ATAX | 4688 | 498928 | 1728 | 936 | 745 | 0 |
| Backprop | 0 | 14282372 | 0 | 1 | 0 | 0 |
| BICG | 8704 | 33797724 | 9952 | 8398 | 8385 | 2224 |
| Hotspot | 6144 | 97340 | 4416 | 31 | 0 | 0 |
| MVT | 2912 | 0 | 2736 | 0 | 0 | 0 |
| NW | 29952 | 8812785 | 23776 | 6651 | 3230 | 772 |
| Pathfinder | 0 | 1878699 | 160 | 7 | 0 | 0 |
| Srad-v2 | 5632 | 44650411 | 5632 | 4209 | 3942 | 3667 |
| 2DCONV | 0 | 566653 | 0 | 0 | 0 | 0 |
| StreamTriad | 0 | 3690578 | 0 | 0 | 0 | 0 |

TABLE VII
TOP-1 ACCURACY OF PAGE DELTA PREDICTION ON MULTIPLE WORKLOADS USING ONLINE TRAINING AND OUR SOLUTION.

| | | Online training | | Our solution | |
|---|---|---|---|---|---|
| | | Streaming | Regular | Streaming | Regular |
| | | 2DCONV | Srad-v2 | 2DCONV | Srad-v2 |
| Streaming | S.T. | 0.908 | 0.772 | 0.874 | 0.80 |
| Regular | Hot. | 0.744 | 0.683 | 0.804 | 0.76 |
| Mixed | NW | 0.560 | 0.478 | 0.849 | 0.78 |
| Random | ATAX | 0.837 | 0.716 | 0.867 | 0.78 |

## VI. RELATED WORKS

This paper is the first to propose an incremental-learning-based approach for CPU-GPU UVM oversubscription management. We now discuss related works in UVM, and studies that apply artificial intelligence to other parts of the microarchitecture.

### A. CPU-GPU UVM Oversubscription Studies

UVM support in modern discrete CPU-GPU systems [37, 38] has been studied widely. Agarwal et al. [39] proposed aggressive first-touch migration and prefetching neighboring pages. Zheng et al. [12] studied different user-directed and user-agnostic prefetchers to overlap data migration and kernel execution. Ganguly et al. [2] uncovered the mechanism of the tree-based prefetcher implemented in the NVIDIA GPU driver. Pratheek et al. [40] proposed walk stealing to reduce interference in page walks from concurrent tenants while also ensuring high walker utilization. Chen et al. [41] proposed an application-transparent framework for reducing memory oversubscription overheads in GPUs. Kim et al. [42] proposed a GPU runtime software and hardware solution that enables efficient demanding paging for GPUs. Ganguly et al. proposed a programmer-agnostic framework [43] and an application-aware adaptive framework [3] to deal with memory oversubscription overhead stemming from page thrashing in irregular, data-intensive GPU applications. Yu et al. [4] proposed a hierarchical page eviction policy that addresses LRU's inability to handle thrashing access patterns while retaining LRU's advantages for LRU-friendly patterns. Yu et al. [44] proposed a coordinated page prefetch and eviction design to manage oversubscription for GPUs with unified memory.

### B. Artificial Intelligence in Computer Architecture

Hashemi et al. [5] apply the RNN model to the analysis of memory access patterns, which demonstrates higher precision and recall than table-based approaches. Peled et al. [18] proposed the context-based prefetcher, which employs the contextual bandits model of reinforcement learning. Bhatia et al. [19] introduced perceptron-based prefetch filtering which acts as an independent check on the quality of predictions made by the underlying prefetch engine. Shi et al. [6] applied deep learning to solve the cache replacement problem. Doudali et al. [45] presented a page scheduler with machine intelligence for applications that execute over hybrid memory systems.

Peled et al. [20] use a fully-connected feed-forward network instead, and they formulate prefetching as a regression problem to train their neural network. Shi et al. [7] propose a hierarchical model of data prefetching that accommodates both delta patterns and addresses correlation. Bera et al. [21] propose a customizable prefetching framework that formulates prefetching as a reinforcement learning problem.

## VII. CONCLUSION

In this paper, we have made a case for a learning-based method to handle memory oversubscription in CPU-GPU UVM. We first provide an in-depth analysis of the current rule-based methods for oversubscription management and the current learning-based methods on other hardware prediction issues. We identify the necessity of applying machine intelligence for accurate data prefetching and eviction to improve the rule-based methods, and we identify sources of performance overhead for the current learning-based method are the explosive growing number of classes and the non-pattern-awareness. Then, we design a framework to solve the problems we found in the analysis. We enable the pattern-awareness of the page predictor by classifying the input data into different categories according to their access patterns, and we train them with a separate neural model. We introduce incremental learning to help the predictor handle the continually growing number of classes. We use a policy engine to deliver prefetching or eviction decisions according to the prediction results. Finally, the evaluation results show that our solution achieves higher performance than the SOTA methods for oversubscription management in CPU-GPU UVM.

Work remains in further improving the page predictor's accuracy. For example, data prefetching can be as accurate as demand load. With the advance in technology, we believe that learning-based methods will become practical and be leveraged by hardware designers. We hope that this paper will inspire the design of other studies which are trying to apply learning-based methods in a heterogeneous systems like CPU-GPU, and multi-GPUs.

## Acknowledgment

TBD